\documentclass[11pt]{article}

\usepackage[margin=1in]{geometry}
\usepackage{amsmath}
\usepackage{amssymb}
\usepackage{graphicx}
\usepackage{booktabs}
\usepackage{siunitx}
\usepackage{subcaption}
\usepackage{float}
\usepackage{caption}
\usepackage[hidelinks]{hyperref}
\usepackage[numbers,sort&compress]{natbib}

\begin{document}

\title{Large-Area SnS Crystals by Controlled Sn--S Chemical Vapor Deposition: Growth Optimization, Morphology, and Raman Characterization}

\author{
Bilal Ahmed$^{1}$\\
{\small $^{1}$Department of Physics, University of South Florida, Tampa, Florida, USA}
}

\date{}
\maketitle

\begin{abstract}
Tin monosulfide (SnS) is a layered IV--VI semiconductor with strong in-plane anisotropy and properties of interest for optoelectronic and ferroic applications. Obtaining large and morphologically well-defined SnS crystals remains challenging. Here, we report the growth of large-area SnS crystals by chemical vapor deposition using separate elemental Sn and S precursors. By varying the precursor temperatures and positions, substrate temperature, carrier-gas flow, and hydrogen concentration, we obtain SnS crystals with lateral dimensions approaching \SI{350}{\micro\meter}. The optimized growth uses a Sn source at approximately \SI{750}{\celsius}, a S source at approximately \SI{230}{\celsius}, a substrate temperature of approximately \SIrange{650}{680}{\celsius}, and Ar/H$_2$ flow rates of 120/10 sccm. X-ray diffraction shows a dominant SnS (111) reflection with no detectable secondary crystalline tin-sulfide phase within the sensitivity of the measurement. Raman spectroscopy reveals six characteristic SnS modes at approximately 39.3, 48.8, 95.2, 165.0, 193.1, and 219.1~cm$^{-1}$. The low-frequency modes near 39 and 49~cm$^{-1}$ are particularly well resolved, with fitted FWHM values of 2.52 and 2.37~cm$^{-1}$, respectively. SEM and AFM measurements show large continuous crystals and a relatively smooth measured surface. Growth on mica produces more regular square and rectangular flakes while retaining large lateral dimensions, with representative flakes approaching approximately $300\times300~\mu\mathrm{m}^{2}$. These results provide a practical growth window for large-area SnS crystals suitable for further optical, electrical, and device studies.
\end{abstract}

\noindent\textbf{Keywords:} SnS; tin monosulfide; chemical vapor deposition; CVD; large-area crystals; Raman spectroscopy; low-frequency phonons; mica; morphology control; two-dimensional materials

\section{Introduction}

Two-dimensional (2D) layered materials have attracted extensive interest because reducing the thickness to the nanoscale can produce distinct electronic, optical, mechanical, and ferroic properties. The development of graphene and transition-metal dichalcogenides (TMDCs), in particular, established mechanically exfoliated and synthetically grown atomically thin crystals as important platforms for studying low-dimensional physics and for developing electronic and optoelectronic devices \cite{Gao2018Exfoliation,Syrgiannis2026}. Tin monosulfide (SnS) is a layered group-IV monochalcogenide with an orthorhombic crystal structure and a composition based on earth-abundant elements. It has attracted interest because of its semiconducting and optical properties, strong in-plane anisotropy, and potential for photovoltaic and other electronic applications \cite{Norton2021,Burton2013}. Optical measurements have reported an indirect fundamental transition near 1.0--1.1~eV and a direct transition near approximately 1.3~eV, although the reported values depend on temperature, polarization, thickness, and measurement conditions \cite{Parenteau1990,Burton2013}. The combination of a suitable optical gap and strong optical absorption has therefore motivated the investigation of SnS for thin-film solar cells and other optoelectronic devices \cite{Sinsermsuksakul2014,Norton2021}.

The low-symmetry crystal structure of SnS also gives rise to strong in-plane anisotropy and unusual ferroic properties. First-principles studies predicted ferroelectric and ferroelastic behavior in monolayer SnS and related group-IV monochalcogenides \cite{Hanakata2016}. Purely in-plane ferroelectricity was subsequently demonstrated experimentally in micrometer-scale SnS at room temperature, showing that these predicted properties can persist in experimentally accessible thin crystals \cite{Higashitarumizu2020}. Other theoretical studies have predicted magnetic behavior associated with suitable substitutional defects or dopants, while persistent spin-helix behavior has also been proposed for group-IV monochalcogenide monolayers including SnS \cite{Ullah2018,Asghar2022,Absor2019}. These results indicate that SnS is not limited to conventional semiconductor applications and may provide a platform for studying coupled electronic, optical, ferroic, and spin-related phenomena.

The layered nature of SnS also makes its thickness an important parameter. Mechanical exfoliation has played a central role in the development of 2D materials, from graphene to semiconducting TMDCs, by providing access to high-quality single- and few-layer crystals from bulk layered precursors \cite{Gao2018Exfoliation,Rashid2019,Ali2024MoS2}. SnS itself has been mechanically thinned to few-layer and ultrathin forms, and monolayer SnS has also been obtained by vapor-phase growth on mica \cite{Hanakata2016,Kawamoto2020}. However, the interlayer interactions in SnS are relatively strong compared with those in some other layered materials because of the lone-pair electrons associated with Sn, making the formation of atomically thin SnS more challenging \cite{Koyama2025,Kawamoto2020}. Consequently, large-area parent crystals can be valuable as a starting platform for subsequent thinning, exfoliation, optical studies, and device fabrication. This consideration is particularly relevant in the broader context of 2D materials, where large-area crystals are desirable for obtaining larger flakes and for reducing the limitations associated with small exfoliated crystal dimensions \cite{Gao2018Exfoliation}.

The potential uses of SnS extend beyond conventional photovoltaic and optoelectronic devices. SnS has been investigated as an absorber material in thin-film solar cells \cite{Sinsermsuksakul2014,Azmy2025Perspective}, and SnS quantum dots have also been demonstrated as inorganic hole-transporting materials in perovskite solar cells \cite{Azmy2025ChemMater}, illustrating the possibility of incorporating SnS into hybrid perovskite architectures \cite{Li2019SnSPerovskite,Azmy2025Perspective}. SnS-based materials have additionally been explored for electrochemical energy-storage applications, including supercapacitor electrodes and related hybrid structures \cite{Sarasamreen2024SnSSupercap}. In photoelectrochemical energy conversion, SnS photocathodes have demonstrated hydrogen-evolution activity, and SnS-based photoelectrodes have been incorporated into bias-free solar water-splitting systems \cite{Lee2021SnSWaterSplitting,Mehmood2018,Bagherifard2025,Kaur2025}. These examples do not imply that large-area CVD SnS crystals are directly optimized for each of these applications; rather, they illustrate the broader range of device and energy-conversion architectures in which SnS and related SnS-based structures are being investigated.

Despite these attractive properties, obtaining large, high-quality, and morphologically well-defined SnS crystals remains challenging. The Sn--S system contains multiple stable and metastable phases, including SnS, Sn$_2$S$_3$, and SnS$_2$, so the relative chemical supply of Sn and S is an important parameter during vapor-phase growth \cite{Clark1999,Koyama2025,Dong2026}. Recent work by Koyama \textit{et al.} demonstrated selective growth of SnS from separate elemental Sn and S precursors by controlling the sulfur vapor concentration and substrate position, and reported SnS crystals with lateral dimensions of approximately \SI{158}{\micro\meter} \cite{Koyama2025}. This work established elemental Sn and S as a useful precursor combination for selective SnS growth, while also highlighting the sensitivity of the material to the growth environment. Other vapor-phase studies have further shown that substrate, temperature, and growth conditions can influence the crystal size, thickness, and morphology of SnS \cite{Lee2019,Kawamoto2020}.

In the present work, we investigate a controlled CVD approach using separate elemental Sn and S precursors to develop a practical growth window for large-area SnS crystals. The effects of precursor temperature and position, substrate temperature, carrier-gas flow, and hydrogen concentration are examined to understand their influence on crystal morphology and lateral size at one atm. Under optimized conditions, we obtain SnS crystals with lateral dimensions approaching \SI{350}{\micro\meter}. We further compare SnS growth on SiO$_2$/Si and mica to evaluate the effect of the substrate on crystal shape and edge definition. Structural and vibrational characterization by X-ray diffraction, Raman spectroscopy, scanning electron microscopy, and atomic force microscopy is used to evaluate the resulting crystals. Particular attention is given to the low-frequency Raman modes, which are clearly resolved in the present samples. The overall objective is therefore not to claim a new device application, but to establish controlled growth conditions for large-area, morphologically well-defined SnS crystals that can serve as a useful materials platform for subsequent optical, electrical, exfoliation, and device studies.
\section{Experimental Methods}

\subsection{CVD Growth}
SnS crystals were synthesized in a horizontal quartz-tube CVD system using separate elemental Sn and S precursors. The Sn precursor was placed in the high-temperature region at approximately \SI{750}{\celsius}, while the sulfur source was positioned in a lower-temperature region at approximately \SI{230}{\celsius}. A relatively larger amount of Sn precursor was used compared with sulfur.

The substrate was positioned downstream from the precursor region. The substrate temperature was varied from approximately \SI{550}{\celsius} to \SI{700}{\celsius}; large crystals were obtained mainly in the \SIrange{650}{680}{\celsius} range. Argon was used at 120 sccm and hydrogen at 10 sccm under the optimized condition. Additional experiments varied the carrier-gas flow and hydrogen concentration to evaluate their effects on crystal morphology and phase formation.

\begin{figure}[H]
    \centering
    \includegraphics[width=0.80\linewidth]{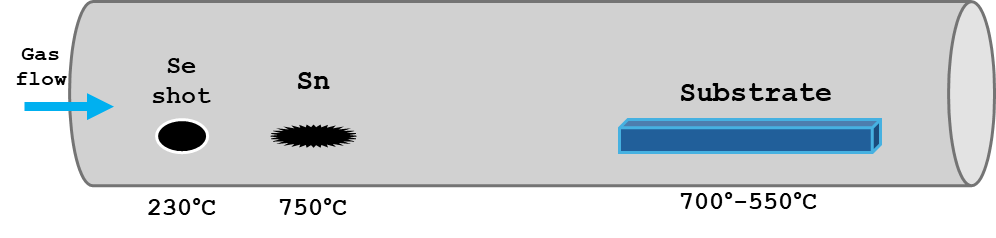}
    \caption{Schematic of the CVD configuration used for the growth of large-area SnS crystals. Separate elemental Sn and S sources are positioned in different temperature regions, with the substrate located downstream from the precursor sources.}
    \label{fig:cvd}
\end{figure}

\section{Results and Discussion}

\subsection{CVD Growth Strategy and Morphology Control}

The use of separate Sn and S sources provides independent control over the supply of the two constituent elements. This is important in the Sn--S system because changes in the relative sulfur concentration can shift the growth between different tin-sulfide phases \cite{Clark1999,Koyama2025,Dong2026}.

The Sn source was maintained at approximately \SI{750}{\celsius}, while the sulfur source was placed near \SI{230}{\celsius}. The substrate temperature was varied between approximately \SI{550}{\celsius} and \SI{700}{\celsius}. The largest lateral crystals were obtained when the substrate temperature was approximately \SIrange{650}{680}{\celsius}. As the substrate temperature decreased toward \SI{550}{\celsius}, the crystals became smaller. This behavior is consistent with the importance of growth temperature in controlling SnS nucleation and morphology reported in vapor-phase growth studies \cite{Lee2019}.

The carrier-gas flow also influenced the growth morphology. Under the optimized condition, Ar and H$_2$ were supplied at 120 and 10 sccm, respectively. Increasing the carrier-gas flow produced more elongated or rod-like structures in our experiments, whereas lower-flow conditions favored a more continuous film. These observations indicate that changes in precursor transport conditions affect the resulting growth morphology.

Hydrogen concentration also affected both morphology and phase formation. Without H$_2$, the deposited material consisted mainly of small rectangular SnS flakes. A moderate H$_2$ flow promoted the formation of larger crystals under the optimized temperature and transport conditions. Increasing the H$_2$ concentration further produced more irregular surfaces and visible defects; at sufficiently high H$_2$ flow, features consistent with SnS$_2$ were also observed. The sensitivity of the Sn--S system to the relative chemical environment is consistent with the phase-selective growth reported for SnS and SnS$_2$ \cite{Clark1999,Koyama2025}.

Under the optimized conditions, consisting of a Sn source at approximately \SI{750}{\celsius}, a sulfur source at approximately \SI{230}{\celsius}, a substrate temperature of approximately \SIrange{650}{680}{\celsius}, and Ar/H$_2$ flow rates of 120/10 sccm, large-area SnS crystals with lateral dimensions approaching \SI{350}{\micro\meter} were obtained. This is larger than the approximately \SI{158}{\micro\meter} crystal reported by Koyama \textit{et al.} using elemental Sn and S precursors on SiO$_2$/Si \cite{Koyama2025}.

\subsection{Structural Characterization of Large-Area SnS}

\begin{figure}[H]
    \centering
    \includegraphics[width=0.68\linewidth]{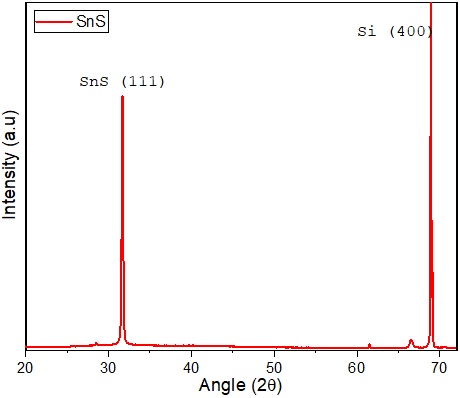}
    \caption{X-ray diffraction pattern of the as-grown SnS crystals. The strong reflection near $2\theta\approx32^\circ$ is assigned to the SnS (111) plane, while the intense feature near $69^\circ$ originates from the Si substrate.}
    \label{fig:xrd}
\end{figure}

The XRD pattern in Figure~\ref{fig:xrd} contains a strong diffraction peak near $2\theta=32^\circ$, consistent with the (111) reflection of orthorhombic SnS \cite{Burton2013,Lee2019}. A second strong reflection near $2\theta=69^\circ$ is attributed to the Si substrate and is consistent with the Si (400) reflection for a Si(100) substrate.

No additional diffraction peaks attributable to a secondary crystalline tin-sulfide phase are detected within the sensitivity of the measurement. In particular, no distinct SnS$_2$ reflection is observed. The XRD pattern therefore supports the formation of crystalline SnS without a detectable secondary crystalline tin-sulfide phase in the measured sample.

The dominance of the SnS (111) reflection may indicate preferred orientation. However, because only one strong SnS reflection is resolved, the overall crystallographic texture cannot be established quantitatively from this measurement alone.

\subsection{Raman Characterization and Phonon Response}

Raman scattering from SnS was established in early work by Chandrasekhar \textit{et al.} in 1977, who measured Raman and infrared spectra of SnS and SnSe and assigned the zone-center phonons according to crystal symmetry \cite{Chandrasekhar1977}. Subsequent Raman studies have examined the characteristic modes, thickness dependence, and polarization dependence of SnS \cite{Li2017,Mehta2017}. Recent work has also examined resonant exciton--phonon coupling in layered SnS, highlighting the continuing importance of phonon spectroscopy in this material \cite{Jadczak2025}.

\begin{figure}[H]
    \centering
    \includegraphics[width=0.76\linewidth]{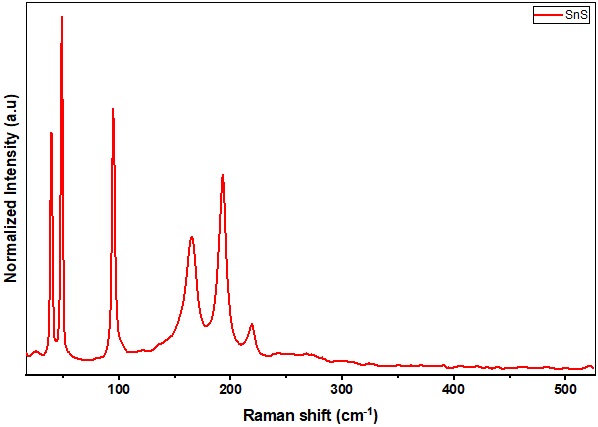}
    \caption{Room-temperature Raman spectrum of the as-grown SnS crystal. Six characteristic SnS modes are observed at 39.3, 48.8, 95.2, 165.0, 193.1, and 219.1~cm$^{-1}$. The low-frequency modes near 39 and 49~cm$^{-1}$ are particularly well resolved.}
    \label{fig:raman}
\end{figure}

The room-temperature Raman spectrum in Figure~\ref{fig:raman} contains six distinct modes at 39.3, 48.8, 95.2, 165.0, 193.1, and 219.1~cm$^{-1}$. These positions agree well with previously reported Raman modes of orthorhombic SnS. Li \textit{et al.} reported six modes at 40.2, 49.1, 95.9, 164.0, 192.0, and 219.5~cm$^{-1}$ in relatively thick SnS flakes \cite{Li2017}. The observed features are therefore assigned to the characteristic $A_g$ and $B_{3g}$ Raman-active modes of SnS.

The most notable feature of the present spectrum is the clarity of the low-frequency response. The $A_g$ mode at 39.3~cm$^{-1}$ and the $B_{3g}$ mode at 48.8~cm$^{-1}$ have fitted FWHM values of 2.52 and 2.37~cm$^{-1}$, respectively. The $A_g$ mode near 95.2~cm$^{-1}$ has a FWHM of 3.71~cm$^{-1}$. These low-frequency modes are clearly resolved in the present measurement.

Several reported SnS spectra emphasize the higher-frequency modes near 190--220~cm$^{-1}$. For example, Mehta \textit{et al.} reported prominent features near 97, 163, 193, and 222~cm$^{-1}$ for CVD-grown SnS \cite{Mehta2017}. Koyama \textit{et al.}, using elemental Sn and S precursors, likewise reported prominent modes in the approximately 94--216~cm$^{-1}$ range \cite{Koyama2025}. In comparison, the present crystals show a strong and distinct response from both low-frequency modes near 39 and 49~cm$^{-1}$ together with the higher-frequency modes.

The observation of all six modes is not claimed as a new observation because the complete six-mode spectrum has been reported previously \cite{Li2017}. The significance of the present Raman result is instead the clear resolution of the low-frequency phonons together with the complete Raman fingerprint. The fitted FWHM values are 2.52, 2.37, 3.71, 21.41, 10.15, and 22.81~cm$^{-1}$ for the modes at 39.3, 48.8, 95.2, 165.0, 193.1, and 219.1~cm$^{-1}$, respectively.

Raman intensity depends on crystal orientation, polarization configuration, excitation wavelength, thickness, absorption, and the Raman tensor elements \cite{Li2017}. Therefore, the strong low-frequency response is not used as a stand-alone measure of crystallinity. Instead, the clearly resolved low-frequency modes, their relatively narrow fitted profiles, the complete six-mode spectrum, and the absence of additional prominent Raman features together provide evidence of a well-defined vibrational response.

No prominent feature is observed near 315~cm$^{-1}$, where the characteristic $A_{1g}$ Raman mode of 2H-SnS$_2$ is commonly reported \cite{Sriv2018}. This is consistent with the absence of a detectable secondary SnS$_2$ phase in the XRD pattern.

\begin{table}[H]
\centering
\caption{Fitted Raman parameters for the six characteristic SnS modes.}
\label{tab:raman}
\begin{tabular}{cccc}
\toprule
Mode & Assignment & Peak position (cm$^{-1}$) & FWHM (cm$^{-1}$) \\
\midrule
1 & $A_g$ & 39.3 & 2.52 \\
2 & $B_{3g}$ & 48.8 & 2.37 \\
3 & $A_g$ & 95.2 & 3.71 \\
4 & $B_{3g}$ & 165.0 & 21.41 \\
5 & $A_g$ & 193.1 & 10.15 \\
6 & $A_g$ & 219.1 & 22.81 \\
\bottomrule
\end{tabular}
\end{table}

\subsection{Morphological and Surface Characterization by SEM and AFM}

\begin{figure}[H]
    \centering
    \begin{subfigure}{0.48\linewidth}
        \centering
        \includegraphics[width=\linewidth]{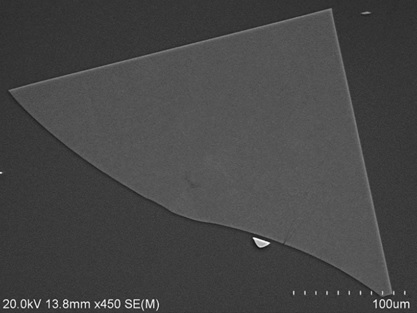}
        \caption{SEM}
    \end{subfigure}
    \hfill
    \begin{subfigure}{0.48\linewidth}
        \centering
        \includegraphics[width=\linewidth]{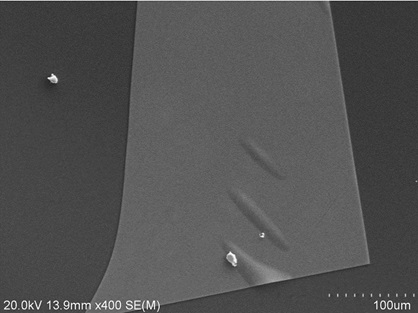}
        \caption{SEM}
    \end{subfigure}

    \vspace{0.2cm}

    \begin{subfigure}{0.48\linewidth}
        \centering
        \includegraphics[width=\linewidth]{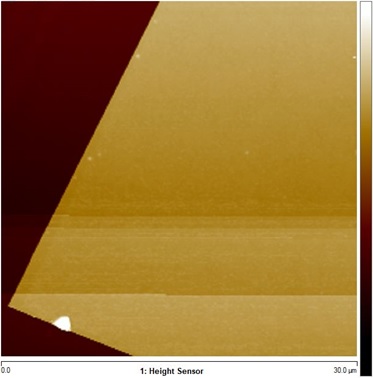}
        \caption{AFM topography}
    \end{subfigure}
    \hfill
    \begin{subfigure}{0.48\linewidth}
        \centering
        \includegraphics[width=\linewidth]{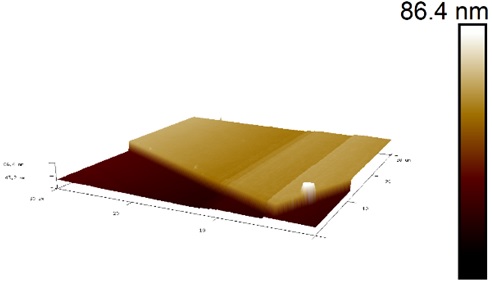}
        \caption{3D AFM}
    \end{subfigure}

    \caption{Morphological and surface characterization of large-area SnS crystals. (a,b) SEM images showing the lateral dimensions and continuous morphology. (c,d) AFM images of a representative surface region showing relatively low nanoscale roughness.}
    \label{fig:sem-afm}
\end{figure}

SEM and AFM were used to evaluate the morphology and nanoscale surface of the large-area crystals. The SEM images in Figure~\ref{fig:sem-afm}(a,b) show broad continuous crystal regions extending over several hundred micrometers. Localized folds, wrinkles, and small particulate features are visible in some regions, but they do not dominate the measured crystal area.

AFM images in Figure~\ref{fig:sem-afm}(c,d) show a relatively smooth surface in the measured region. The average roughness is $R_a\approx0.23$~nm and the root-mean-square roughness is $R_q\approx0.30$~nm, with a maximum height variation of approximately 3.07~nm over the analyzed region. These values describe the measured surface region and are not taken as representative of the entire crystal.

Separate AFM measurements on representative flakes yielded thicknesses on the order of tens of nanometers, including a representative flake approximately 80~nm thick. The AFM data shown in Figure~\ref{fig:sem-afm} are used primarily to characterize surface morphology and roughness rather than to assign the thickness of that particular crystal.

\subsection{Substrate-Dependent Morphology: SiO$_2$/Si and Mica}

\begin{figure}[H]
    \centering
    \begin{subfigure}{0.48\linewidth}
        \centering
        \includegraphics[width=\linewidth]{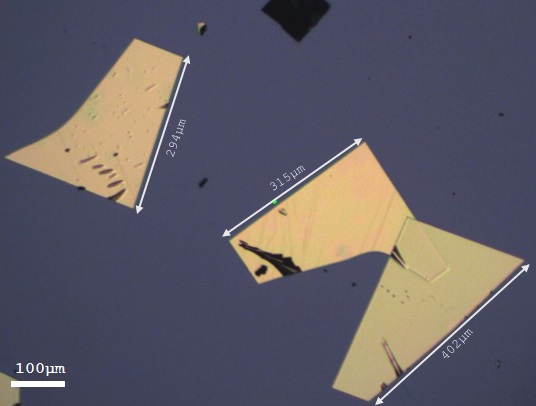}
        \caption{SiO$_2$/Si}
    \end{subfigure}
    \hfill
    \begin{subfigure}{0.48\linewidth}
        \centering
        \includegraphics[width=\linewidth]{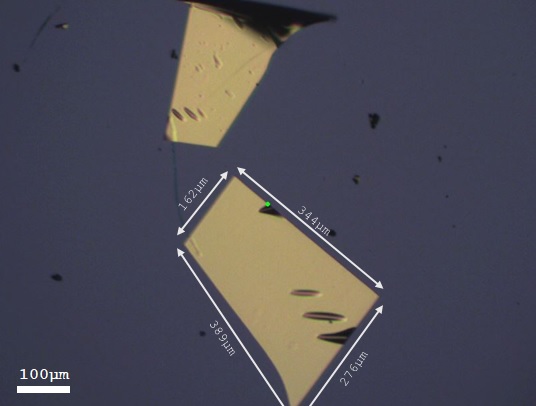}
        \caption{SiO$_2$/Si}
    \end{subfigure}

    \vspace{0.2cm}

    \begin{subfigure}{0.48\linewidth}
        \centering
        \includegraphics[width=\linewidth]{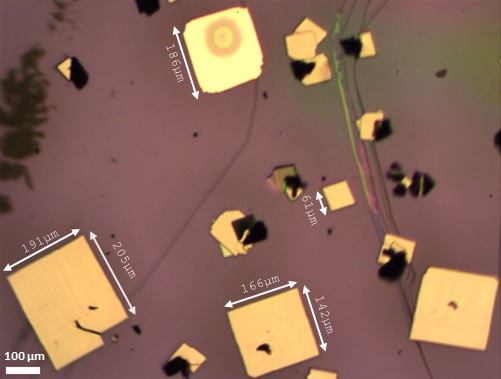}
        \caption{Mica}
    \end{subfigure}
    \hfill
    \begin{subfigure}{0.48\linewidth}
        \centering
        \includegraphics[width=\linewidth]{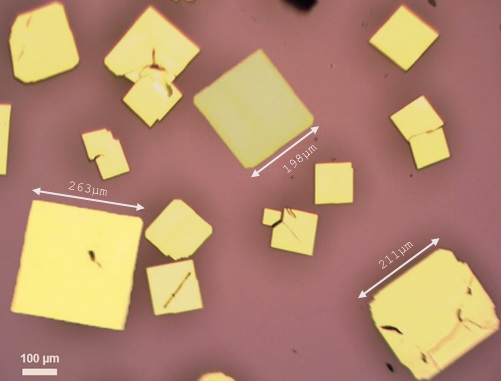}
        \caption{Mica}
    \end{subfigure}

    \caption{Comparison of SnS crystal morphology on SiO$_2$/Si and mica. (a,b) Crystals grown on SiO$_2$/Si showing large lateral dimensions with irregular or fractured edges. (c,d) Crystals grown on mica showing more regular square and rectangular morphologies while retaining large lateral dimensions.}
    \label{fig:substrates}
\end{figure}

Although the optimized growth on SiO$_2$/Si produced crystals with lateral dimensions approaching \SI{350}{\micro\meter}, the crystal edges were not always well defined. As shown in Figure~\ref{fig:substrates}(a,b), crystals on SiO$_2$/Si commonly exhibit irregular or fractured boundaries.

Mica has previously been used as a substrate for SnS growth, including highly textured films, thin films, and micrometer-scale monolayer crystals \cite{Wang2014,Wang2017Mica,Higashitarumizu2020,Kawamoto2020}. Previous studies also show that the mica surface can influence SnS nucleation and crystal orientation \cite{Wang2014,Kawamoto2020}.

The crystals grown on mica in the present work show a more regular square or rectangular morphology with better-defined lateral edges, as shown in Figure~\ref{fig:substrates}(c,d). We do not assign a specific microscopic mechanism for this change because the SnS/mica interface was not directly investigated. Instead, the result is treated as an experimental observation of substrate-dependent morphology.

Importantly, mica does not prevent large lateral growth under the present conditions. Representative square and rectangular flakes approach approximately $300\times300~\mu\mathrm{m}^{2}$. Thus, the significance of the mica result is not the first observation of rectangular SnS crystals, which has been reported previously, but the ability to retain large lateral dimensions while obtaining more regular crystal geometries \cite{Wang2017Mica,Higashitarumizu2020,Kawamoto2020}. Similar substrate-assisted morphology control has also been demonstrated for the related SnSe system \cite{Chiu2023}.

\section{Conclusions}

We demonstrated a CVD growth window for large-area SnS crystals using separate elemental Sn and S precursors. Control of precursor temperature and position, substrate temperature, carrier-gas flow, and hydrogen concentration allowed the growth morphology to be adjusted, with optimized conditions producing crystals with lateral dimensions approaching \SI{402}{\micro\meter}. XRD showed a dominant SnS (111) reflection and no detectable secondary crystalline tin-sulfide phase within the measurement sensitivity.

Raman spectroscopy showed the six characteristic SnS modes at 39.3, 48.8, 95.2, 165.0, 193.1, and 219.1~cm$^{-1}$. The low-frequency modes near 39 and 49~cm$^{-1}$ were particularly well resolved, with fitted FWHM values of 2.52 and 2.37~cm$^{-1}$. Together with the XRD, SEM, and AFM results, the Raman data support a well-defined crystalline and vibrational response of the grown material.

Comparison of substrates showed that SiO$_2$/Si supports very large crystals but can produce irregular or fractured edges, whereas mica produced more regular square and rectangular flakes while retaining large lateral dimensions, with representative flakes approaching $300\times300~\mu\mathrm{m}^{2}$. The resulting large-area crystals provide a useful platform for future polarization-dependent spectroscopy, electrical characterization, and device fabrication.

\section*{Data Availability}
The data that support the findings of this study are available from the corresponding author upon reasonable request.

\bibliographystyle{unsrtnat}
\bibliography{references}

\end{document}